\documentclass[letterpaper,10pt,twocolumn]{article}
\usepackage{graphics}
\usepackage{amssymb}
\usepackage{graphics}
\usepackage{cite}

\begin{document}

\title{An overview of ordered adlayer structures for CO and NO\\ on the (100) surface of Pt, Rh, Ni, Cu and Pd}
\author{C. G. M. Hermse\thanks{email: chretien@sg10.chem.tue.nl}\\
\small{Schuit Institute of Catalysis, ST/SKA, Eindhoven University of Technology, P. O. Box 513, 5600 MB Eindhoven, The Netherlands}\\
}

\maketitle

\newpage
\begin{abstract}
Many ordered adlayer structures have been reported for CO on the (100) surfaces of Pt, Rh, Pd, Ni and Cu. Also, some ordered structures are known for NO on the same surfaces. The current review includes all those articles that give a graphical representation of the position of the adsorbates inside the unit cell. The emphasis is on the case of CO on Pt(100) and Rh(100), but also structures are included for adsorption on the (100) surface of other transition metals, as well as references for the case of NO adsorption. 
\end{abstract}

\section*{Results}

\begin{figure}
\includegraphics{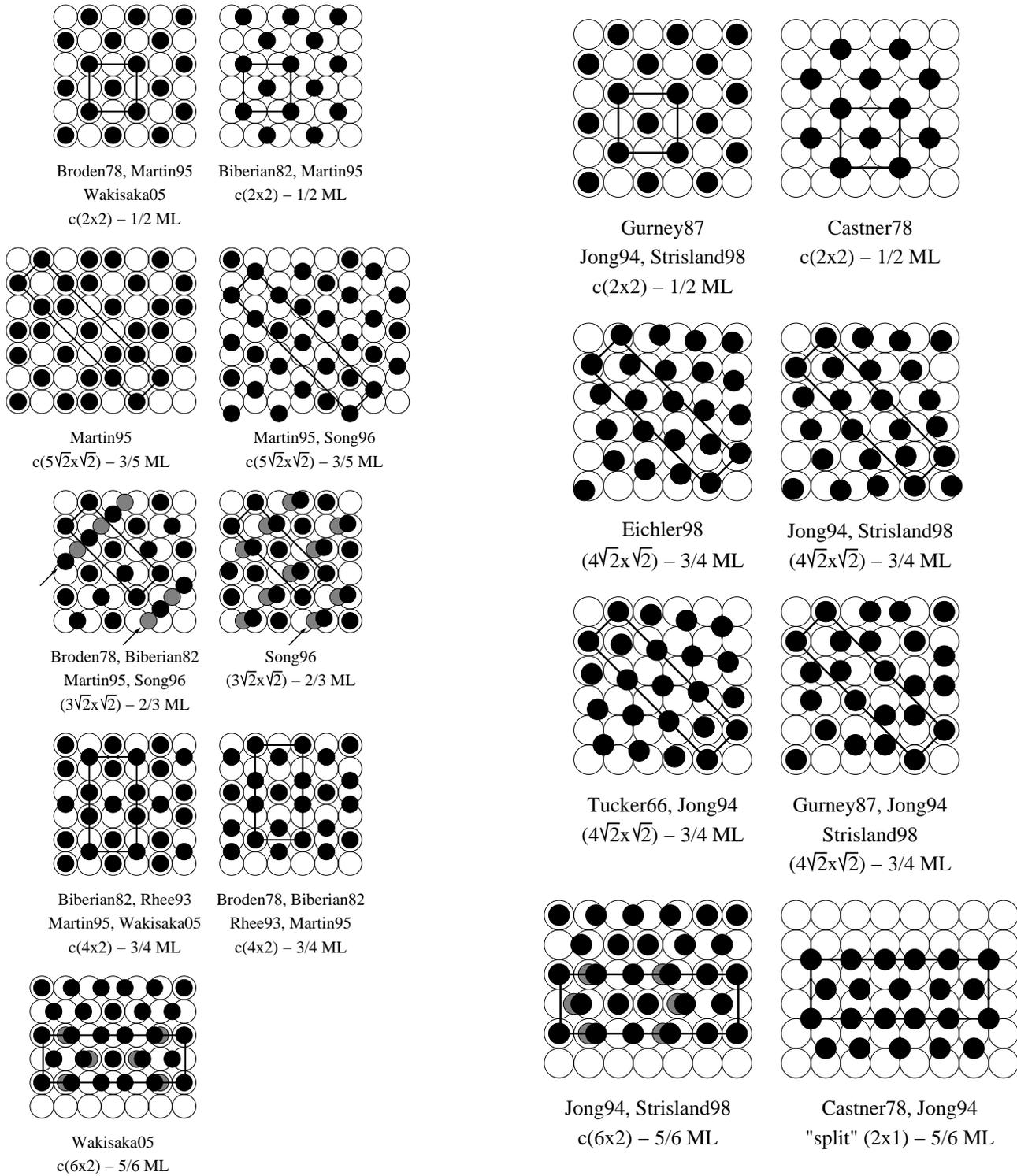}
\caption{Ordered structures for CO on the Pt(100) surface.\cite{mar95,bib82,son96,bro78,rhe93,wak05} The black circles indicate the positions of the CO molecules as reported in experimental literature, the grey circles indicate the alternative positions of the CO molecules according to a recent Monte Carlo model \cite{her08pre}.
\label{co_pt100}}
\end{figure}

\begin{figure}
\includegraphics{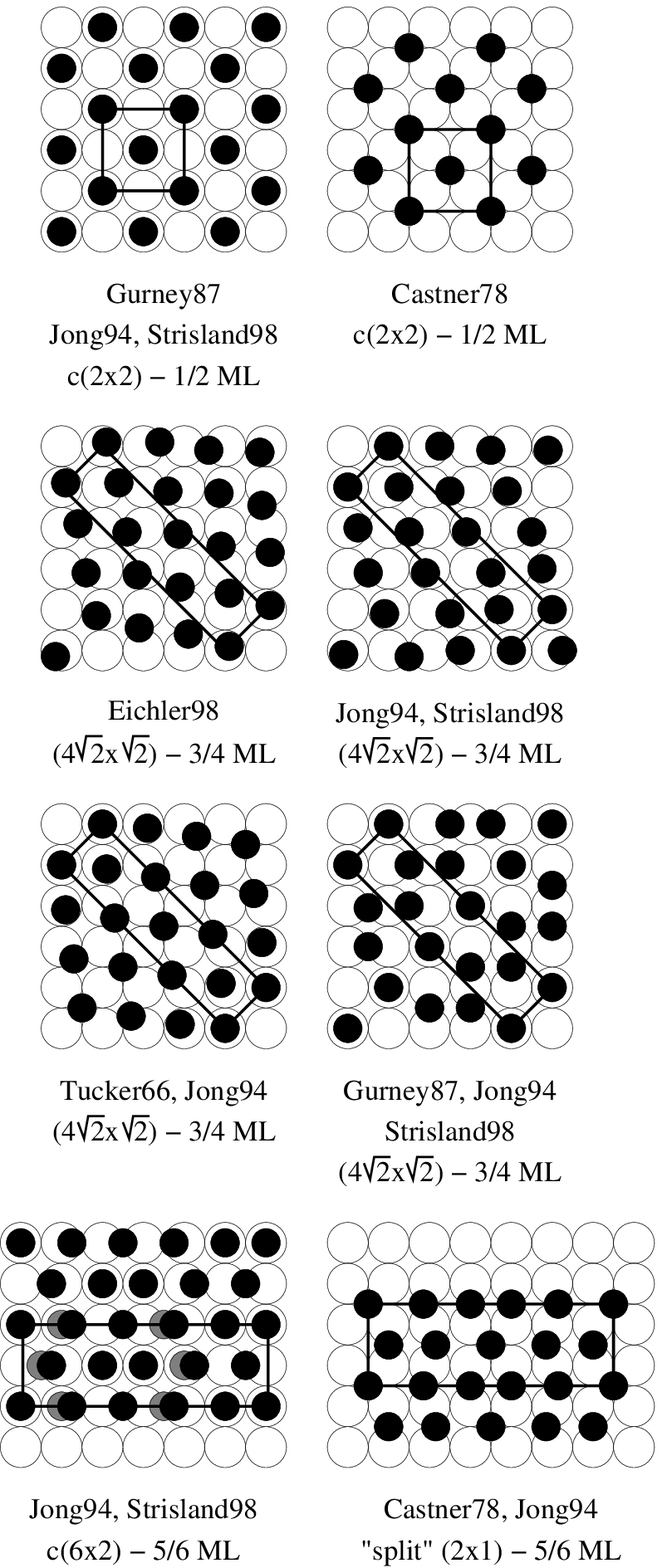}
\caption{Ordered structures for CO on the Rh(100) surface \cite{jong,jon94,str98,gur87,cas78,eic98,tuc66}. The black circles indicate the positions of the CO molecules as reported in experimental literature, the grey circles indicate the alternative positions of the CO molecules according to a recent Monte Carlo model \cite{her08pre}.
\label{co_rh100}}
\end{figure}

\begin{figure}
\includegraphics{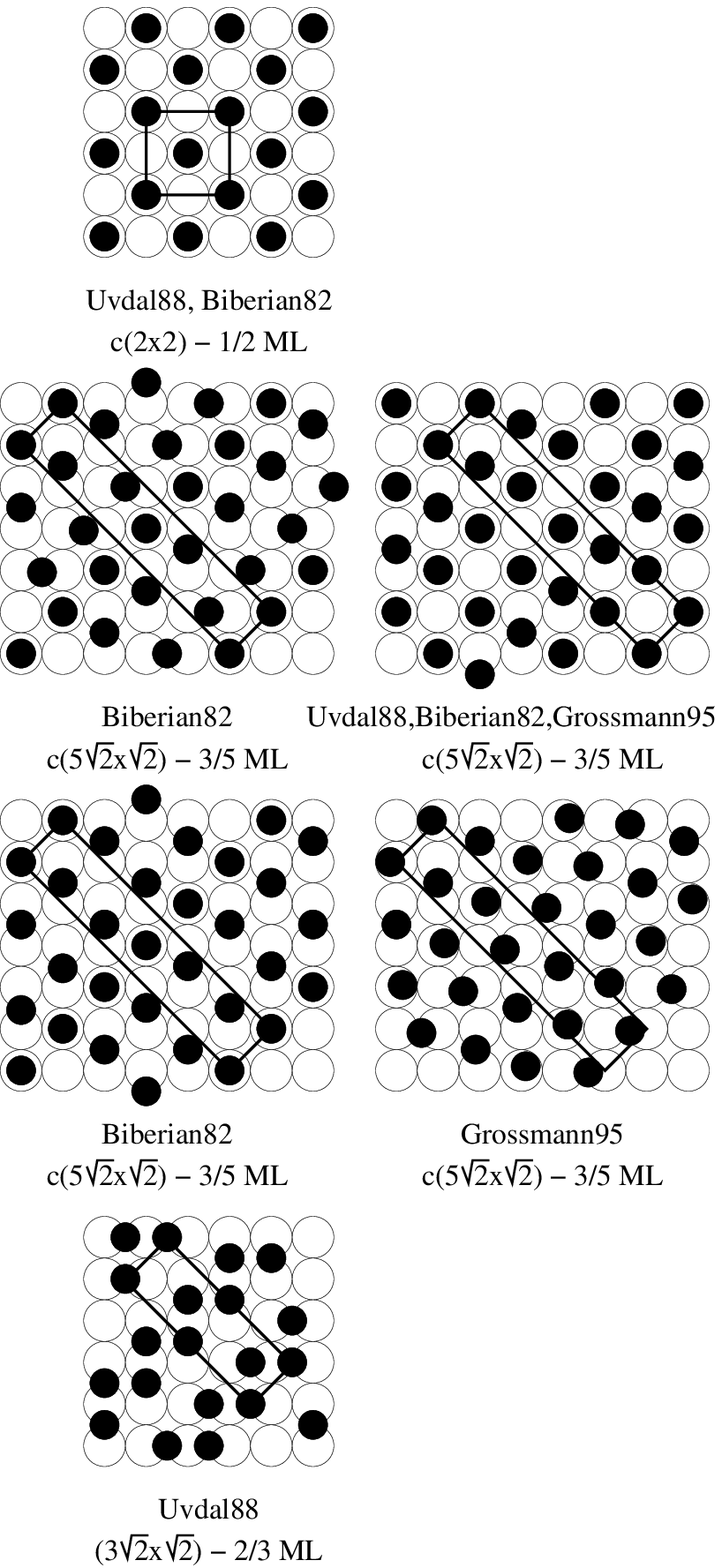}
\caption{Ordered structures for CO on the Ni(100) surface.\cite{uvd88,bib82,gro95} 
\label{co_ni100}}
\end{figure}

\begin{figure}
\includegraphics{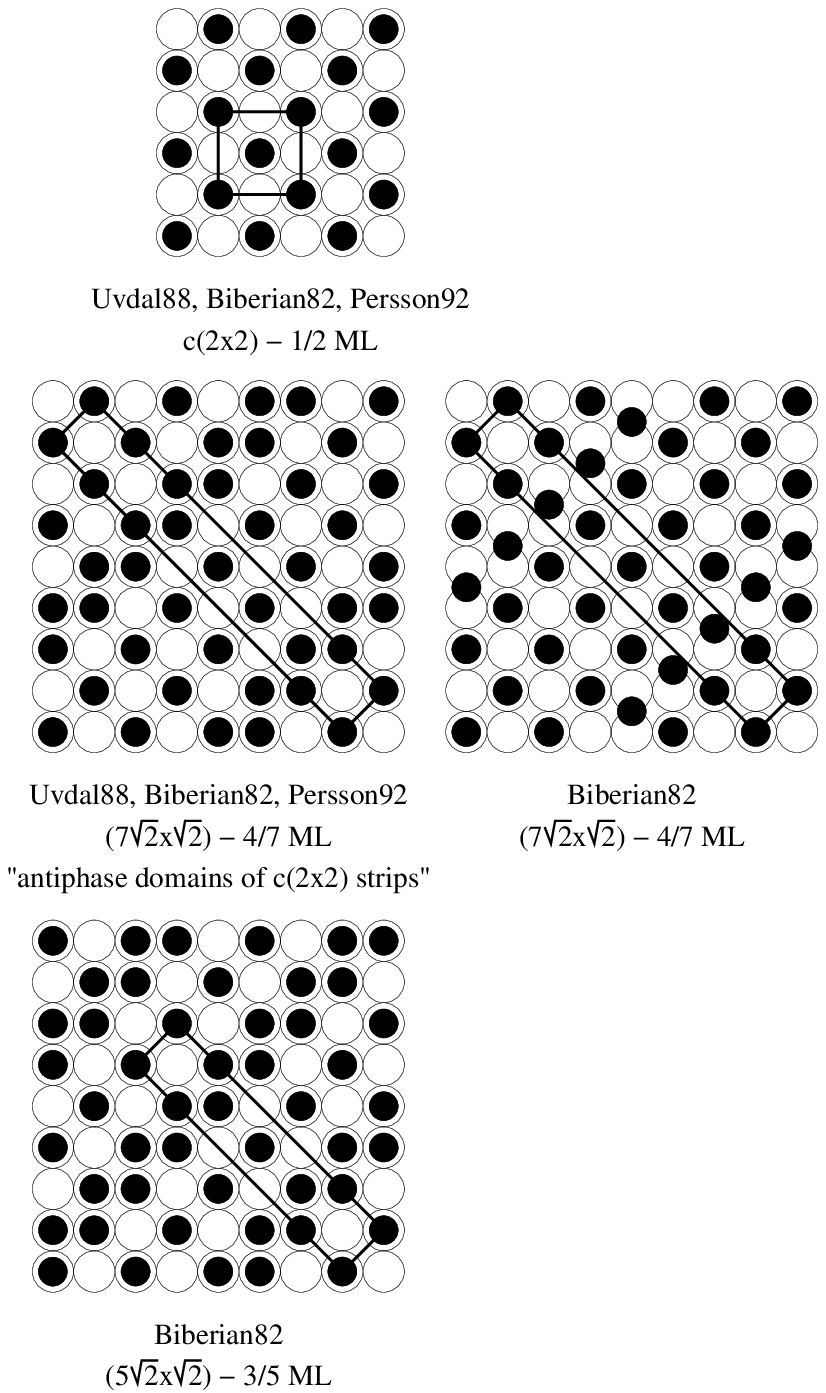}
\caption{Ordered structures for CO on the Cu(100) surface.\cite{bib82,uvd88,per92}. 
\label{co_cu100}}
\end{figure}

\begin{figure}
\includegraphics{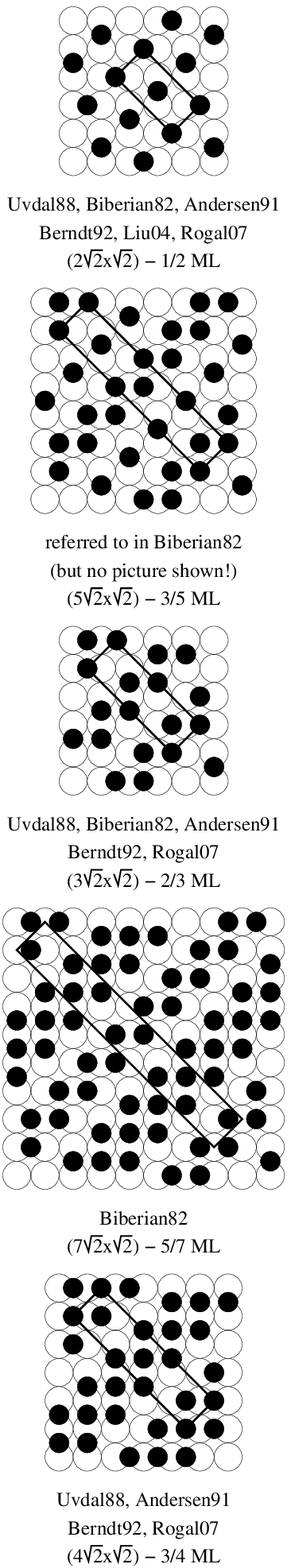}
\caption{Ordered structures for CO on the Pd(100) surface.\cite{bib82,uvd88,and91,ber92,liu04,rog07} In addition to these structures, domain wall superlattices are cited in \cite{ber92}.
\label{co_pd100}}
\end{figure}

\begin{figure}
\includegraphics{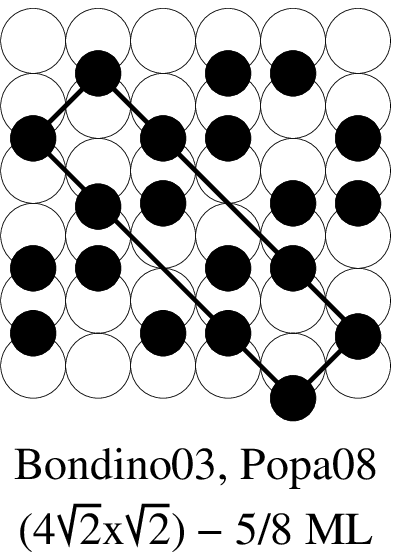}
\caption{Ordered structure for NO on the Rh(100) surface.\cite{bon03,bon03a,pop08}
\label{no_rh100}}
\end{figure}

The figures show for each adsorbate / transition metal surface combination the ordered structures that were reported, and in which paper they were reported. The indication of the paper consists of the last name of the first author, combined with the last two digits of the year it was published in. The full reference is listed in the References section. 

\subsection*{Ordered structures for CO}

Representations of ordered structures of CO on Pt(100), Rh(100), Ni(100), Cu(100) and Pd(100) can be found in the Figures 1--5.

\subsection*{Ordered structures for NO}
The representation of the ordered structure of NO on Rh(100) can also be found in Figure 6. 

Representations of ordered structures of NO on Pt(100) can be found in \cite{son00,son97,bon78,gar90a,eic01,her06}. 

No representations of ordered structures of NO on Ni(100) were found during this review. No representations of ordered structures of NO on Cu(100) were found during this review. 

Representations of ordered structures of NO on Pd(100) can be found in \cite{jaw02}.

\section*{Concluding remarks}
Comments or suggestions for improvement of this document are welcome. Additional references are appreciated.

\section*{Version}
June 11th, 2009, version 1.0.





\end{document}